\documentclass[a4,11pt]{article}
\usepackage{amsmath,amssymb}
\usepackage[dvipdfmx]{graphicx}
\begin{document}

\title{Contact dynamics, contact Poisson bracket, and symplectic integrator 
-- Even Arnold nodds}
\author{Kiyoshi \textsc{Sogo} 
\thanks{EMail: sogo@icfd.co.jp}\ and Shuhei \textsc{Ohnishi}
}

\maketitle

\begin{center}
Institute of Computational Fluid Dynamics, 1-16-5, Haramachi, Meguroku, 
Tokyo, 152-0011, Japan
\end{center}

\abstract{
By introducing an integration factor to the differential one-form of contact dynamics, equations of motion are 
derived variationally, and contact Poisson bracket and contact Lagrangian are formulated. 
Discrete symplectic integrator, named hybrid leap-frog method, is found to be a numerical solver of equations 
of motion for contact dissipative systems. 
}
%
%
\section{Introduction}
\setcounter{equation}{0}

Theory of symplectic geometry is a mathematical analogue of the canonical theory of classical mechanics. \cite{Arnold}
Since the total number of variables is even, such as $2n$ of $p=(p_1,\cdots, p_n),\ q=(q_1,\cdots,q_n)$ in classical mechanics, 
the symplectic space (or phase space) has usually even dimensions. 
Fundamental object of symplectic dynamics is the differential one-form
\begin{align}
\omega=pdq-H(p, q)dt
\end{align}
which is usually a differential $dS$ of the action function $S$ in Hamilton-Jacobi theory.
Equations of motion are derived by the variational principle for 
$S=\int\omega$,
\begin{align}
&\delta S=\int \left[\delta p\left( \frac{dq}{dt}-\frac{\partial H}{\partial p}\right)-
\delta q\left(\frac{dp}{dt}+\frac{\partial H}{\partial q}\right)\right]dt=0
\nonumber \\
&\Longrightarrow\quad
\frac{dq}{dt}=\frac{\partial H}{\partial p},\quad 
\frac{dp}{dt}=-\frac{\partial H}{\partial q}.
\end{align}

On the other hand, the contact geometry \cite{Arnold} \cite{Bravetti} \cite{deLeon} 
is an analogue of symplectic geometry for {\it odd} ($2n+1$) dimensional space, 
whose differential one-form is given by ($n=1$)
\begin{align}
\omega=pdq+dz-K(p, q, z)dt,
\label{contactK}
\end{align}
where $K$ is called contact Hamiltonian, and the motion driven by $K$ is called {\it contact dynamics}. 
The integrability condition of \eqref{contactK} can be derived by using Frobenius theorem \cite{CartanH}
\begin{align}
\Omega=\sum_{i=1}^N P_i(x)\ dx_i\ &\text{is integrable} 
\nonumber\\
&\text{if and only if}\ 
\sum_{(i,j,k)} \epsilon_{ijk}\ P_i\ \frac{\partial P_k}{\partial x_j}=0,
\label{Frobenius}
\end{align}
where $\epsilon_{ijk}$ is the Levi-Civita symbol ({\it i.e.} completely anti-symmetric tensor) 
with three numbers $(i, j, k)$ chosen from $N$ numbers. 
There are simultaneous conditions of totally $\binom{N}{3}$ combinations in general.
Now by setting $(x_1, x_2, x_3; P_1, P_2, P_3)=(q, z, t; p, 1, -K)$ in \eqref{Frobenius}, and regarding 
$\partial p/\partial t=dp/dt$ in heuristic manner, we obtain
\begin{align}
\frac{dp}{dt}=-\frac{\partial K}{\partial q}+p\frac{\partial K}{\partial z},
\end{align}
which is one of the equations of motion for contact dynamics to be derived in the next section. 
It should be noted the last term in the right hand side is a new term not existing in ordinary canonical dynamics. 

The first problem of contact dynamics is how to obtain all equations of motion, especially for $z$ variable 
which has apparently no conjugate momentum. 
The key idea to the answer is to introduce an {\it integration factor} $\lambda$, 
which might be a conjugate momentum of $z$ variable.
Since the history of contact geometry is very old \cite{Lie},
there are certainly many literatures discussing this problem, {\it e.g.} 
the appendix 4 of Arnold's book. \cite{Arnold} Unfortunately it does not give the correct equation of motion 
for $\lambda$,  
\begin{align}
\text{Arnold's}\quad \dot{\lambda}=-K_z\quad \text{should be}\quad \dot{\lambda}=-\lambda K_z,
\label{Arnold}
\end{align}
and does not give general formula for contact version of Poisson bracket $\{A, B\}$ for arbitrary $A, B$ pair. 
Although Arnold employed the former of \eqref{Arnold} by setting $\lambda=1$, 
he should have introduced $\mu=\log\lambda$ by writing $\dot{\mu}=-K_z$ instead. 
In other words, the variable $\lambda$ has an aggressive role behind the whole theory of contact dynamics, 
as will be shown later. 
All equations of motion and general contact Poisson bracket will be derived later in {\it 2.1} and {\it 2.2} 
respectively. 
Within the authors' knowledge,  there are no satisfactory discussions in literatures on these problems.  
Contact Lagrangian and its relation with Herglotz equation is discussed in {\it 2.3}. 

The aim of the present paper is to fully discuss the {\it symplectic} structure of contact dynamics, 
such as the equations of motion for $p, q, z$ and $\lambda$, and general expression of the contact Poisson bracket 
which contains $\lambda$-differential terms. 
Symplectic integrator is introduced in {\it 3.1}, and is applied in {\it 3.2} to several examples which contain {\it dissipations}. 
This possibility to treat dissipative systems with mathematical consistency is our physical motivation to consider 
the contact geometry.

\section{Contact dynamics, contact Hamiltonian, and contact Poisson bracket}
\setcounter{equation}{0}
\subsection{Contact dynamics driven by contact Hamiltonian}

Let us introduce an integration factor $\lambda$ to our one-form $\omega=pdq+dz-Kdt$, 
then we have
\begin{align}
\begin{split}
\Omega&=\lambda\omega=\lambda pdq+\lambda dz-\lambda K(p,q,z)dt
\\
&\equiv p_1dq_1+p_0dq_0-H(p_1,q_1,p_0,q_0) dt,
\end{split}
\label{oneform}
\end{align}
where we set
\begin{align}
H(p_1=\lambda p, q_1=q, p_0=\lambda, q_0=z)=\lambda\ K(p, q, z).
\end{align}
Because such $\Omega$ is an ordinary even dimensional symplectic one-form, 
it gives Hamilton's equations of motion 
\begin{align}
\frac{dq_1}{dt}=\frac{\partial H}{\partial p_1},\quad
\frac{dp_1}{dt}=-\frac{\partial H}{\partial q_1},\quad
\frac{dq_0}{dt}=\frac{\partial H}{\partial p_0},\quad
\frac{dp_0}{dt}=-\frac{\partial H}{\partial q_0}.
\label{Hamilton}
\end{align}

Since variables relations $p_1=\lambda p,\ p_0=\lambda$ is written inversely 
as $p=p_1/p_0,\ \lambda=p_0$, we have
\begin{align}
\begin{split}
&\frac{\partial}{\partial p_1}=\frac{\partial p}{\partial p_1}\frac{\partial}{\partial p}+
\frac{\partial \lambda}{\partial p_1}\frac{\partial}{\partial\lambda}=\frac{1}{\lambda}\frac{\partial}{\partial p},
\\
&\frac{\partial}{\partial p_0}=\frac{\partial p}{\partial p_0}\frac{\partial}{\partial p}+
\frac{\partial \lambda}{\partial p_0}\frac{\partial}{\partial\lambda}=
-\frac{p}{\lambda}\frac{\partial}{\partial p}+\frac{\partial}{\partial\lambda},
\end{split}
\label{key}
\end{align}
therefore we can rewrite by substituting $H=\lambda K(p, q, z)$ such as
\begin{align}
\begin{split}
&\frac{\partial H}{\partial p_1}=\frac{1}{\lambda}\frac{\partial (\lambda K)}{\partial p}=\frac{\partial K}{\partial p},
\\
&\frac{\partial H}{\partial q_1}=\frac{\partial(\lambda K)}{\partial q}=\lambda\frac{\partial K}{\partial q},
\\
&\frac{\partial H}{\partial p_0}=\frac{\partial(\lambda K)}{\partial\lambda}-
\frac{p}{\lambda}\frac{\partial(\lambda K)}{\partial p}=K-p\frac{\partial K}{\partial p},
\\
&\frac{\partial H}{\partial q_0}=\frac{\partial(\lambda K)}{\partial z}=\lambda\frac{\partial K}{\partial z}.
\end{split}
\end{align}
Then Hamilton's equations \eqref{Hamilton} become
\begin{align}
\begin{split}
&\frac{dq}{dt}=\frac{\partial K}{\partial p},\quad
\frac{d(\lambda p)}{dt}=-\lambda\frac{\partial K}{\partial q},
\\
&\frac{dz}{dt}=K-p\frac{\partial K}{\partial p},\quad
\frac{d\lambda}{dt}=-\lambda\frac{\partial K}{\partial z},
\end{split}
\end{align}
from which we obtain the final results,
\begin{align}
\dot{q}=K_p,\quad \dot{p}=-K_q+pK_z,\quad \dot{z}=K-pK_p,\quad \dot{\lambda}=-\lambda K_z,
\label{EoMfinal}
\end{align}
where notations such as $\dot{q}=dq/dt$ and $K_p=\partial K/\partial p$, and equalities
\begin{align}
\frac{d(\lambda p)}{dt}=\lambda\frac{dp}{dt}+p\frac{d\lambda}{dt}=\lambda\dot{p}-p\lambda K_z,
\end{align}
are used. 
It is remarkable that the first three equations in \eqref{EoMfinal} contain no variable $\lambda$, 
and only the last equation $\dot{\lambda}=-\lambda K_z$ contains $\lambda$.

For the sake of clarity, let us give another derivation of \eqref{EoMfinal}, 
which is essentially same but seemingly different, by applying variational principle to
\begin{align}
S=\int\left(\lambda pdq+\lambda dz-\lambda Kdt\right)
=\int\lambda\left(p\frac{dq}{dt}+\frac{dz}{dt}-K\right)dt.
\end{align}
Taking the variation of $S$, we have the integrand of $\delta S$ such that
\begin{align}
&\delta\lambda\left(p\frac{dq}{dt}+\frac{dz}{dt}-K\right)+
\delta p\left(\lambda\frac{dq}{dt}-\lambda K_p \right)
\nonumber \\
&-\delta q\left(\frac{d(\lambda p)}{dt}+\lambda K_q\right)-
\delta z\left(\frac{d\lambda}{dt}+\lambda K_z\right),
\nonumber
\end{align}
from which we have
\begin{align}
p\frac{dq}{dt}+\frac{dz}{dt}=K,\quad
\frac{dq}{dt}=K_p,\quad
\frac{d(\lambda p)}{dt}=-\lambda K_q,\quad
\frac{d\lambda}{dt}=-\lambda K_z.
\end{align}
After recombining these as before, we obtain 
\begin{align}
\dot{z}=K-pK_p,\quad 
\dot{q}=K_p,\quad
\dot{p}=-K_q+pK_z,\quad 
\dot{\lambda}=-\lambda K_z,
\end{align}
which are \eqref{EoMfinal}. 
It should be emphasized that variation by $\lambda$ derives the equation $\dot{z}=K-p\dot{q}=K-pK_p$, combined with 
a result of other variation. 

It should be remarked here about {\it time inversion symmetry} of variables, 
which usually depends also on the Hamiltonian itself. 
If we write time inversion operation by $T$ such as $T(t)=-t$, what are the behavior of variables $(p, q, \lambda, z)$ ? 
For ordinary dynamics it is usual that Hamiltonian and positions are $T$-even and momenta are $T$-odd. 
However such rule not necessarily holds for our contact dynamics.  
Under several assumptions we can check it by examining equations of motion. 
From \eqref{EoMfinal} we find $T(p, q, \lambda, z)=(-p, q, \lambda, -z)$ under the assumptions 
that covariant equations and invariant $K$. 
Therefore, contrary to intuition, it is reasonable to assume that coordinate $z$ is $T$-odd and momentum $\lambda$ is $T$-even.
We must be careful however that Hamiltonian $K$ is not necessarily $T$-invariant. 
For an example, Hamiltonian
\begin{align}
K=\frac{1}{2}\left(p^2+\omega^2q^2\right)-\gamma z,
\end{align}
is {\it not} $T$-invariant (the first two are $T$-even but the last is $T$-odd), and it describes damped harmonic oscillator, 
whose equation is neither $T$-covariant. 
We will return to this model in {\it 3.2}. 

It should be commented here about the work by the first author, \cite{SogoDissipation}
which gives another (completely different) variational derivation of dissipative equations, 
which include damping equation, Bloch equation, diffusion equation and Fokker-Planck equations.

\subsection{Contact Poisson bracket}

Our system has two Hamiltonians $K$ and $H=\lambda K$, and by use of equations of motion \eqref{EoMfinal} we have
\begin{align}
\frac{dK}{dt}&=\dot{p}\ \frac{\partial K}{\partial p}+\dot{q}\ \frac{\partial K}{\partial q}+
\dot{z}\ \frac{\partial K}{\partial z}
\nonumber \\
&=\left(-K_q+pK_z\right)K_p+K_pK_q+\left(K-pK_p\right)K_z
=KK_z,\\
\frac{dH}{dt}&=\frac{d(\lambda K)}{dt}=\dot{\lambda}\ K+\lambda\ \dot{K}
=\left(-\lambda K_z\right)K+\lambda KK_z=0,
\end{align}
that is, the conserved Hamiltonian is not $K$ but $H$. 
Therefore the Poisson bracket should be defined using $(p_1, q_1, p_0, q_0)=(\lambda p, q, \lambda, z)$ by
\begin{align}
\{A,\ B\}=\sum_{j=0}^1\left(\frac{\partial A}{\partial q_j}\frac{\partial B}{\partial p_j}-
\frac{\partial B}{\partial q_j}\frac{\partial A}{\partial p_j}\right).
\label{Poisson0}
\end{align}
We rewrite this in terms of $(p, q, \lambda, z)$, by using the relations
\begin{align}
\frac{\partial}{\partial q_0}=\frac{\partial}{\partial z},\quad 
\frac{\partial}{\partial p_0}=\frac{\partial}{\partial \lambda}-\frac{p}{\lambda}\frac{\partial}{\partial p},\quad
\frac{\partial}{\partial q_1}=\frac{\partial}{\partial q},\quad
\frac{\partial}{\partial p_1}=\frac{1}{\lambda}\frac{\partial}{\partial p},
\end{align}
where some non-trivial equalities are already derived in \eqref{key}. 
Therefore \eqref{Poisson0} is rewritten as
\begin{align}
&\{A, B\}=
\frac{1}{\lambda}\left(\frac{\partial A}{\partial q}\frac{\partial B}{\partial p}-
\frac{\partial B}{\partial q}\frac{\partial A}{\partial p}\right)
\nonumber \\
&\qquad\qquad +\left\{\frac{\partial A}{\partial z}
\left(\frac{\partial B}{\partial\lambda}-\frac{p}{\lambda}\frac{\partial B}{\partial p}\right)-
\frac{\partial B}{\partial z}
\left(\frac{\partial A}{\partial\lambda}-\frac{p}{\lambda}\frac{\partial A}{\partial p}\right)\right\}.
\end{align}
Let us define here
\begin{align}
\{A, B\}_\text{c}&\equiv\lambda\cdot \{A, B\}
\nonumber \\
&=\left(\frac{\partial A}{\partial q}\frac{\partial B}{\partial p}-
\frac{\partial B}{\partial q}\frac{\partial A}{\partial p}\right)
\nonumber \\
&\quad+\left[\frac{\partial A}{\partial z}
\left(\lambda\frac{\partial B}{\partial\lambda}-p\frac{\partial B}{\partial p}\right)-
\frac{\partial B}{\partial z}
\left(\lambda\frac{\partial A}{\partial\lambda}-p\frac{\partial A}{\partial p}\right)\right],
\label{Poisson3}
\end{align}
and call this {\it contact Poisson bracket} hereafter. 
We should be aware of $\lambda$-differential terms, which is a unique feature of our bracket. 

We can check the bracket \eqref{Poisson3} satisfies 
the basic properties,
\begin{align}
\{A, B\}_\text{c}=-\{B, A\}_\text{c},\quad
\{A, BC\}_\text{c}=\{A, B\}_\text{c}\ C+B\ \{A, C\}_\text{c}.
\label{Property}
\end{align}

The time development of arbitrary variable $A(q, p, z, \lambda)$ is given by
\begin{align}
\frac{dA}{dt}&=\{A, H\}=\frac{1}{\lambda}\ \{A, H\}_\text{c}=\frac{1}{\lambda}\ \{A, \lambda K\}_\text{c}
\nonumber \\
&=\{A, K\}_\text{c}+\frac{1}{\lambda}\{A, \lambda\}_\text{c} K=
\{A, K\}_\text{c}+A_zK,
\end{align}
where by definition $\{A, H\}_\text{c}=\lambda\cdot\{A, H\}$ (put $B=H$) and properties \eqref{Property} are used. 
The last equality $\{A, \lambda\}_\text{c}=\lambda\ A_z$ is derived by subtituting $B=\lambda$ in \eqref{Poisson3}. 

Let us write our final result again
\begin{align}
\frac{dA}{dt}=\{A, K\}_\text{c}+A_zK,
\label{contactPoisson}
\end{align}
which implies that for $z$-independent $A$, the right hand side is $\{A, K\}_\text{c}$ (the usual form), 
and for $z$-dependent $A$, the right hand side has additional term $A_zK$. 
Equation \eqref{contactPoisson} is our Hamilton's equation of motion for contact dynamics. 

For the confirmations, let us verify equations of motion by using \eqref{contactPoisson}. 
Substituting $A=q$ and $A=p$, we have respectively
\begin{align}
\begin{split}
&\frac{dq}{dt}=\{q, K\}_\text{c}+(q)_zK=K_p,\\
&\frac{dp}{dt}=\{p, K\}_\text{c}+(p)_zK=-K_q+pK_z,
\end{split}
\end{align}
where \eqref{Poisson3} is used. 
And for $A=z$ and $A=\lambda$, we have respectively
\begin{align}
\begin{split}
&\frac{dz}{dt}=\{z, K\}_\text{c}+(z)_zK=-pK_p+K,
\\
&\frac{d\lambda}{dt}=\{\lambda, K\}_\text{c}+(\lambda)_zK=-\lambda K_z,
\end{split}
\end{align}
where we used $\{z, K\}_\text{c}=-pK_p$, and 
$\{\lambda, K\}_\text{c}=-\lambda K_z$ by direct substitution into \eqref{Poisson3}.  
These equations are same as \eqref{EoMfinal}.

\subsection{Contact Lagrangian and Herglotz equation}

For ordinary symplectic systems, Lagrangian $L(q, \dot{q})$ and Hamiltonian $H(p, q)$ has a relation
\begin{align}
L(q, \dot{q})+H(p, q)=p\dot{q}, \qquad 
p=\frac{\partial L}{\partial\dot{q}},\quad \dot{q}=\frac{\partial H}{\partial p}.
\label{LHrelation}
\end{align}
Solving $p=\partial L/\partial\dot{q}$ for $\dot{q}$ as $\dot{q}=\dot{q}(p, q)$ we obtain $H(p, q)$ from $L(q, \dot{q})$ 
by using \eqref{LHrelation}. Inversely solving $\dot{q}=\partial H/\partial p$ for $p$ as $p=p(q, \dot{q})$ 
we obtain $L(q, \dot{q})$ from $H(p, q)$. 
If such procedures are not possible, we call them {\it singular} Lagrangian or {\it singular} Hamiltonian respectively. 
The theory of electro-magnetic field is a famous example of such singular cases. 

For the contact dynamics, such $L$-$H$ relation might be analogously assumed as
\begin{align}
L(q, \dot{q},z,\dot{z})+\lambda K(p, q, z)=\lambda p\dot{q}+\lambda\dot{z}.
\label{contactLHrelation0}
\end{align}
This $L$ turns out however to be not suitable because $\lambda$ cannot be eliminated. 
We may instead introduce $J$ by setting $L=\lambda J$, such as  
\begin{align}
J(q, \dot{q},z,\dot{z})+K(p, q, z)=p\dot{q}+\dot{z},
\label{contactJKrelation1}
\end{align}
which becomes however
\begin{align}
J=p\dot{q}-pK_p,
\label{contactJKrelation3}
\end{align}
by using $\dot{z}=K-pK_p$, the third equation of  \eqref{EoMfinal}. 
Then such $J$ is still singular because conjugate momentum $p_z=\partial J/\partial\dot{z}=0$, 
because $\dot{z}$ is absent in \eqref{contactJKrelation3}. 

Thus we finally arrive at a surmise for the suitable $J$-$K$ relation
\begin{align}
J(q,\dot{q},z)=p\dot{q}-K(p.q.z),\qquad
p=\frac{\partial J}{\partial\dot{q}},\quad
\dot{q}=\frac{\partial K}{\partial p}.
\label{contactJKrelation}
\end{align}
It should be noted that \eqref{contactJKrelation} still lacks $\dot{z}$ variable, 
and no way to derive $\dot{\lambda}=-\lambda K_z$, which implies our $J$ remains singular.
The extended Euler-Lagrange equation for $J$ of \eqref{contactJKrelation} will be given by
\begin{align}
\frac{d}{dt}\left(\frac{\partial J}{\partial\dot{q}}\right)-\frac{\partial J}{\partial q}
+\frac{\partial J}{\partial\dot{q}}\frac{\partial J}{\partial z}=0,
\label{Herglotz}
\end{align}
because it is rewritten, by using \eqref{contactJKrelation}, as
\begin{align}
\frac{dp}{dt}=-K_q+pK_z,
\end{align}
which is the second equation of \eqref{EoMfinal}. 
The equation \eqref{Herglotz} is called Herglotz equation. \cite{Herglotz} 
Such relationship with Herglotz equation confirms our $J$-$K$ relation \eqref{contactJKrelation} is suitable one 
for the contact dynamics. We can therefore call $J$ of \eqref{contactJKrelation} the contact Lagrangian  
corresponding to the contact Hamiltonian $K$. 
It should be noted however that to derive the rest equation of \eqref{EoMfinal} $\dot{z}=K-pK_p$ we must 
assume further an equality $\dot{z}+J=0$, since the right hand side of \eqref{contactJKrelation} turns out 
to be $pK_p-K$. Such equality can be expressed as
\begin{align}
dz=-J dt\quad\Longrightarrow\quad z=-\int J(q, \dot{q}, z)\ dt,
\label{HerglotzAction}
\end{align}
which implies variable $z$ is (minus of) the action integral. 
It is the situation of Herglotz variational principle to derive \eqref{Herglotz} 
that variable $z$ appears in both sides of \eqref{HerglotzAction}.

\section{Symplectic integrator for the contact dynamics}
\setcounter{equation}{0}
\subsection{Hybrid leap-frog method}

Let us start from Hamilton's equations of motion \eqref{Hamilton} 
\begin{align}
\frac{dq_0}{dt}=\frac{\partial H}{\partial p_0},\quad
\frac{dp_0}{dt}=-\frac{\partial H}{\partial q_0},\quad
\frac{dq_1}{dt}=\frac{\partial H}{\partial p_1},\quad
\frac{dp_1}{dt}=-\frac{\partial H}{\partial q_1},
\end{align}
which are discretized with time step $h$ by the leap-frog method as
\begin{align}
&p_0^{n+1/2}=p_0^n-\frac{h}{2}\left(\frac{\partial H}{\partial q_0}\right)^n,\ 
q_0^{n+1}=q_0^n+h\left(\frac{\partial H}{\partial p_0}\right)^{n+1/2},
\nonumber \\ 
&p_0^{n+1}=p_0^{n+1/2}-\frac{h}{2}\left(\frac{\partial H}{\partial q_0}\right)^{n+1},
\nonumber \\
&p_1^{n+1/2}=p_1^n-\frac{h}{2}\left(\frac{\partial H}{\partial q_1}\right)^n,\  
q_1^{n+1}=q_1^n+h\left(\frac{\partial H}{\partial p_1}\right)^{n+1/2},
\nonumber \\ 
&p_1^{n+1}=p_1^{n+1/2}-\frac{h}{2}\left(\frac{\partial H}{\partial q_1}\right)^{n+1},
\nonumber
\end{align}
where the super-index $n$ implis discrete time $t=nh$.

It should be noted that the leap-frog method is one of the symplectic integrators which conserve 
the canonical property, \cite{SogoUno2011} such as the Poisson bracket and the volume element, 
{\it i.e.} Liouville theorem. We should comment here on the work by Vermeeren {\it et.al.}, 
\cite{Vermeeren} which is based on different principle, and gives a different integrator from ours. 
Whether their scheme is symplectic or not, is not clarified enough unfortunately.

In original variables, above leap-frog equations are rewritten by
\begin{align}
&\lambda^{n+1/2}=\lambda^n-\frac{h}{2}\left(\lambda K_z\right)^n,\quad  
z^{n+1}=z^n+h\left(K-pK_p\right)^{n+1/2},
\nonumber \\ 
&\lambda^{n+1}=\lambda^{n+1/2}-\frac{h}{2}\left(\lambda K_z\right)^{n+1},\nonumber \\
&(\lambda p)^{n+1/2}=(\lambda p)^n-\frac{h}{2}(\lambda K_q)^{n},\quad  
q^{n+1}=q^n+h(K_p)^{n+1/2},\nonumber \\ 
&(\lambda p)^{n+1}=(\lambda p)^{n+1/2}-\frac{h}{2}(\lambda K_q)^{n+1},\nonumber
\end{align}
which are rearranged as a sequential procedure as follows
\begin{align}
\begin{split}
&\lambda^{n+1/2}=\lambda^n\cdot\left(1-\frac{h}{2} K_z^n\right),\\
&p^{n+1/2}=\left(\lambda^n/\lambda^{n+1/2}\right)\cdot\left(p^n-\frac{h}{2} K_q^n\right),
\\
&z^{n+1}=z^n+h\left(K-pK_p\right)^{n+1/2},\\ 
&q^{n+1}=q^n+h\left( K_p \right)^{n+1/2},\\
&\lambda^{n+1}=\lambda^{n+1/2}/\left(1+\frac{h}{2}K_z^{n+1}\right),\\
&p^{n+1}=\left(\lambda^{n+1/2}/\lambda^{n+1}\right)p^{n+1/2}-\frac{h}{2} K_q^{n+1}.
\end{split}
\end{align}
This procedure is an explicit solver almost everywhere, except the place $(K-pK_p)^{n+1/2}$ and $(K_p)^{n+1/2}$, because 
Hamiltonian $K$ may contain variables not computed at the time $n+1/2$ in general. 
To remedy such defect we can append further  
\begin{align}
z^{n+1/2}=z^n+\frac{h}{2}\left(K-pK_p\right)^n,\quad 
q^{n+1/2}=q^n+\frac{h}{2}\left(K_p\right)^n,
\end{align}
which make all computations explicit. 
Therefore our discretization may be called {\it hybrid leap-frog} method, 
since it is a mixed version of two leap-frog methods
\begin{align}
&(1)\quad p^{n+1/2}=p^n-\frac{h}{2}H_q^n,\quad q^{n+1}=q^n+hH_p^{n+1/2},
\nonumber \\ 
&\qquad p^{n+1}=p^{n+1/2}-\frac{h}{2}H_q^{n+1/2},\\
&(2)\quad q^{n+1/2}=q^n+\frac{h}{2}H_p^n,\quad p^{n+1}=p^n-hH_q^{n+1/2},
\nonumber \\ 
&\qquad q^{n+1}=q^{n+1/2}+\frac{h}{2}H_p^{n+1/2}.
\end{align}

It should be noted that the difference equation of $p^n$ is coupled with motions of $\lambda^n$'s, 
which shall become apparently decoupled by eliminating them such that
\begin{align}
\begin{split}
&p^{n+1/2}=\left(p^n-\frac{h}{2}K_q^n\right)/\left(1-\frac{h}{2}K_z^n\right),
\\
&p^{n+1}=p^{n+1/2}\left(1+\frac{h}{2}K_z^{n+1}\right)-\frac{h}{2}K_q^{n+1},
\end{split}
\end{align}
which are however both non-trivial discretization of $\dot{p}=-K_q+pK_z$ that we could not have imagined without 
mediation of $\lambda^n$'s. 

We will discuss in the next subsection several numerical examples using this hybrid leap-frog method.
 
\subsection{Numerical examples}

\noindent
(A) Damped harmonic oscillator I

Let us consider a contact Hamiltonian 
\begin{align}
K=\frac{1}{2}\left( p^2+\omega^2q^2\right)-\gamma z,
\end{align}
which is not invariant under the time inversion, as mentioned before, since 
$T(q, p, z, \lambda)=(q, -p, -z, \lambda)$. 
Equations of motion are given by
\begin{align}
\begin{split}
&\dot{q}=K_p=p,\quad
\dot{p}=-K_q+pK_z=-\omega^2q-\gamma p,\\
&\dot{z}=K-pK_p=\frac{1}{2}\left(\omega^2q^2-p^2\right)-\gamma z,\quad
\dot{\lambda}=-\lambda K_z=+\gamma \lambda,
\end{split}
\end{align}
from which we have
\begin{align}
\dot{q}=p,\quad \dot{p}=-\omega^2q-\gamma p
\quad\Longrightarrow\quad 
\ddot{q}+\gamma\dot{q}+\omega^2 q=0,
\label{LdHO}
\end{align}
which is a damped harmonic oscillator. And the rest equations are
\begin{align}
\dot{z}+\gamma z=\frac{1}{2}\left(\omega^2 q^2-p^2\right),\quad 
\dot{\lambda}-\gamma\lambda=0,
\end{align}
which can be solved by using the solution of \eqref{LdHO}. \\

\noindent
(B) Damped harmonic oscillator II

Let us consider next the Hamiltonian with nonlinear dependence on $z$
\begin{align}
K=\frac{1}{2}\left( p^2+\omega^2q^2\right)-\gamma z^2,
\end{align}
which is {\it invariant} under the time inversion $T(q, p, z, \lambda)=(q, -p, -z, \lambda)$. 
Equations of motion are
\begin{align}
\begin{split}
&\dot{q}=p,\quad
\dot{p}=-\omega^2q-2\gamma zp,\\
&\dot{z}=\frac{1}{2}\left(\omega^2q^2-p^2\right)-\gamma z^2,\quad
\dot{\lambda}=+2\gamma z\lambda,
\end{split}
\label{NonlinearDecay}
\end{align}
from which we obtain
\begin{align}
\ddot{q}+2\gamma z\dot{q}+\omega^2q=0,
\end{align}
where the damping coefficient depends on $z$.

Contrary to the previous example (A), \eqref{NonlinearDecay} are not solvable analytically.  
It is easy however to apply the hybrid leap-frog integrator to these cases (A) and (B). 
Figure 1 shows numerical results for both cases with time step $h=0.01$ and parameters $\omega=1.0,\ \gamma=0.1$ 
under the initial condition $q(0)=z(0)=\lambda(0)=1.0$ and $p(0)=0.0$. 
The case (B) has a slower decay than the case (A) with the same values of parameters.

\begin{figure}[h]
\centering
\includegraphics[width=8cm, clip]{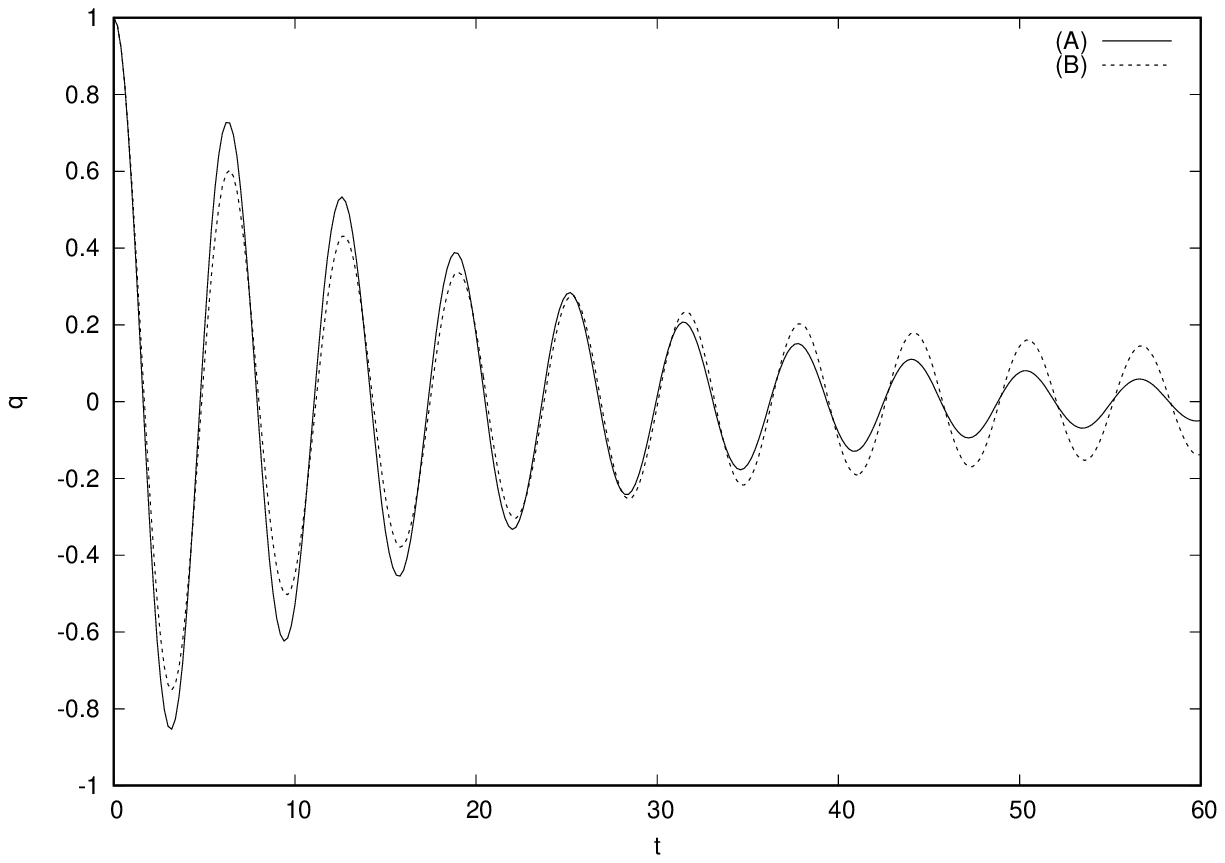}
\caption{$q(t)$ of damped harmonic oscillators I (A) and II (B).}
\end{figure}

\newpage

\noindent
(C) Damped double-well potential

Let us consider next a contact Hamiltonian with nonlinear potential
\begin{align}
K=\frac{p^2}{2}+U(q) - \gamma z,\quad
U(q)=\frac{1}{2}(q^2-a^2)^2,
\end{align}
whose equations of motion are
\begin{align}
\begin{split}
&\dot{q}=p,\quad \dot{p}=-2q(q^2-a^2)-\gamma p,\\
&\dot{z}=\frac{1}{2}(q^2-a^2)^2-\frac{p^2}{2}-\gamma z,\quad
\dot{\lambda}=+\gamma\lambda.
\end{split}
\label{DoublePotential}
\end{align}
Potential $U(q)$ is called {\it double-well potential} having two minimum points at $q=\pm a$. 

Figure 2 shows numerical results for time step $h=0.01$, with parameters $a=1.0,\ \gamma=0.1$,  
initial conditions $z(0)=\lambda(0)=1.0,\ p(0)=0.0$ 
and $q(0)=\pm 2.0$. The results show that the space inversion symmetry is satisfied, 
although the time reversal symmetry is broken due to dissipations.

\begin{figure}[h]
\centering
\includegraphics[width=8cm, clip]{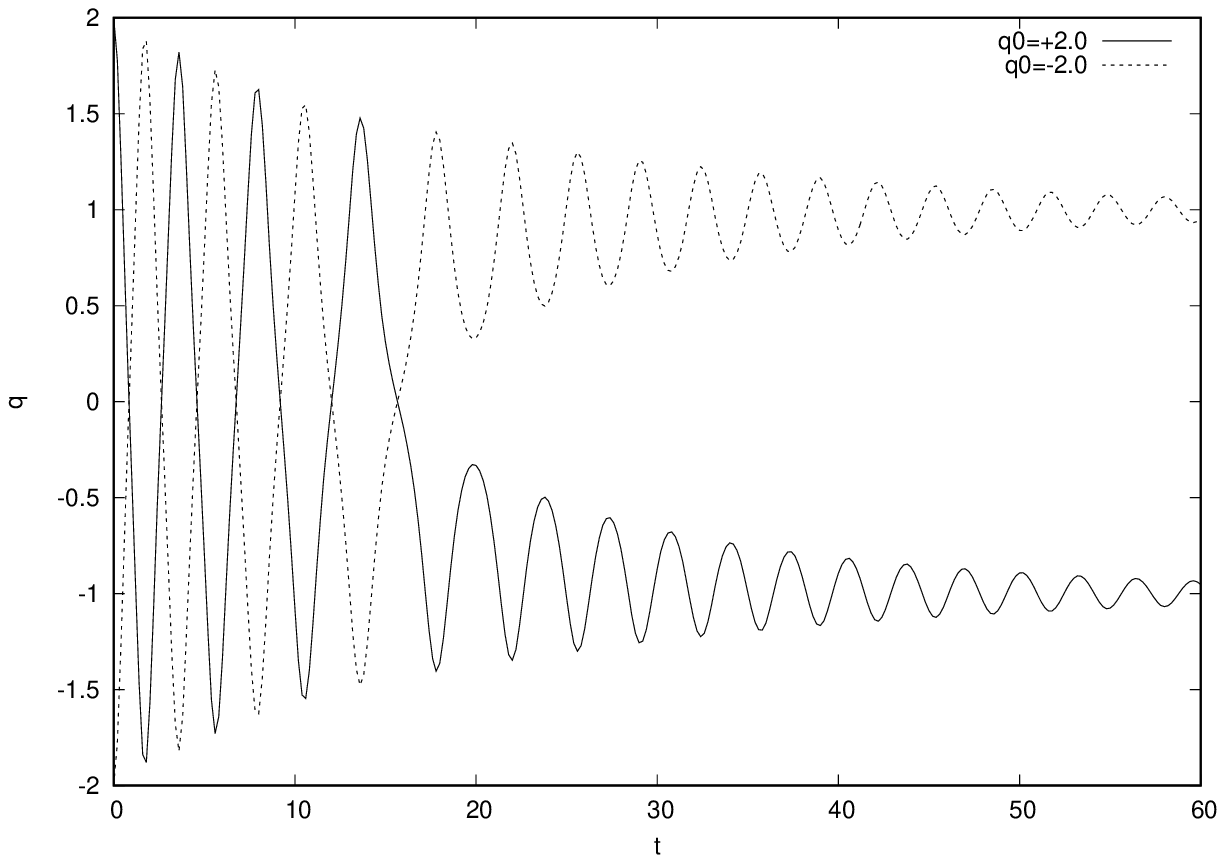}
\caption{$q(t)$ of damped double-well potential under two initial conditions.}
\end{figure}

\newpage

\noindent
(D) Synchronization of two coupled oscillators

Let us consider finally a contact Hamiltonian 
\begin{align}
K=\frac{1}{2}\left(p_1^2+\omega_1^2q_1^2\right)+\frac{1}{2}\left(p_2^2+\omega_2^2q_2^2\right)+gq_1^2q_2^2
-\gamma z,
\end{align}
whose equations of motion are
\begin{align}
\begin{split}
&\dot{q}_1=p_1,\quad \dot{p}_1=-\omega_1^2q_1-2gq_1q_2^2-\gamma p_1,\\
&\dot{q}_2=p_2,\quad \dot{p}_2=-\omega_2^2q_2-2gq_1^2q_2-\gamma p_2,\\
&\dot{z}=\frac{1}{2}\left(\omega_1^2q_1^2-p_1^2\right)+\frac{1}{2}\left(\omega_2^2q_2^2-p_2^2\right)+gq_1^2q_2^2
-\gamma z,\quad  
\dot{\lambda}=+\gamma\lambda.
\end{split}
\end{align}
Even when $\omega_1\neq \omega_2$, sufficiently large coupling $g$ induces two oscillators a synchronous motion. 
Parameters are taken $h=0.01,\ \gamma=0.01,\ \omega_1^2=1.2,\ \omega_2^2=0.8$. 
Initial conditions are $z(0)=\lambda(0)=1.0,\ p_1(0)=p_2(0)=0.0,\ q_1(0)=1.0,\ q_2(0)=-1.0$. 
The left of Figure 3 shows $q_1(t), q_2(t)$ for $g=0.0$ (no coupling), which move independently without synchronization. 
On the contrary, the right of Figure 3 shows behaviors for $g=0.8$, which are in synchronization. 

It should be remarked that if $g=0.8$ and $\gamma=0$, two oscillators make a synchronous motion without decay, 
{\it i.e.} ordinary canonical dynamics. 
  
\begin{figure}[h]
\centering
\includegraphics[width=7.5cm, clip]{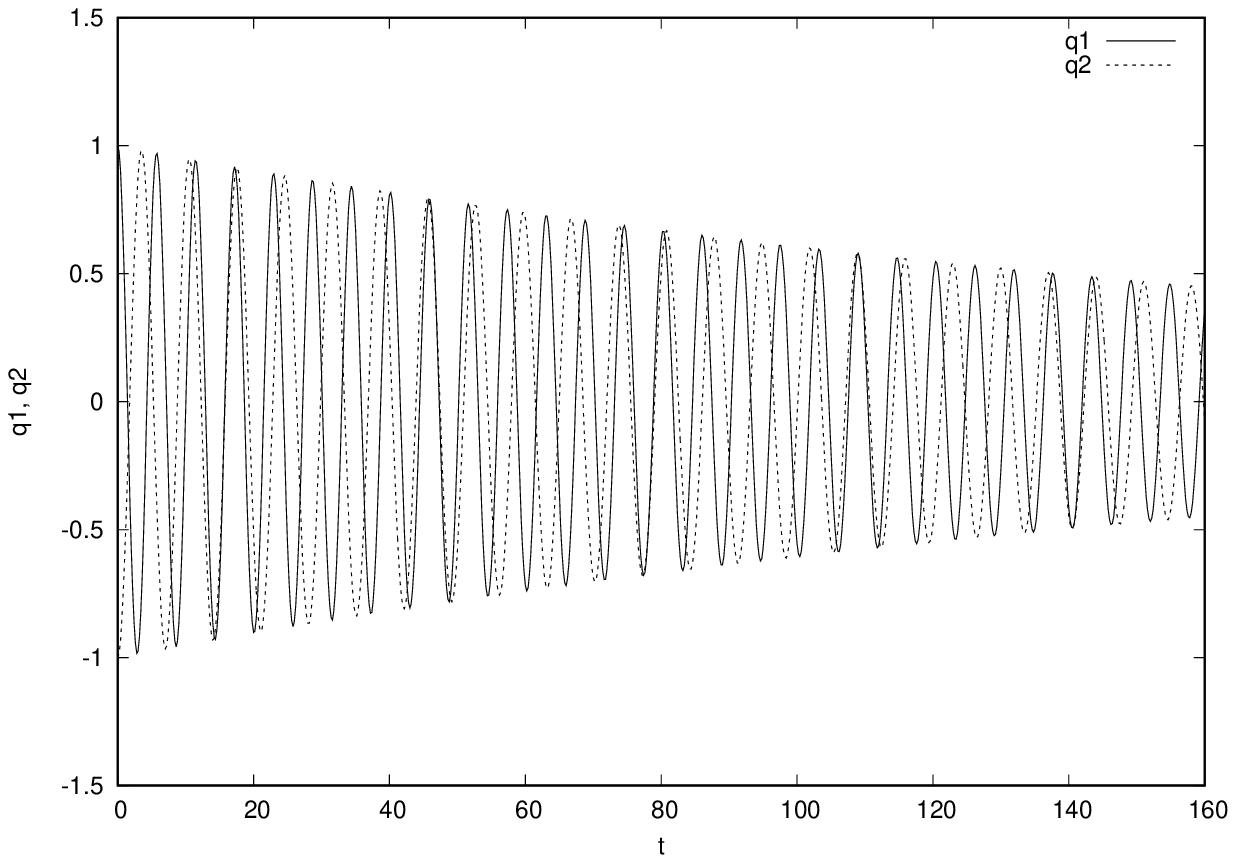}
\includegraphics[width=7.5cm, clip]{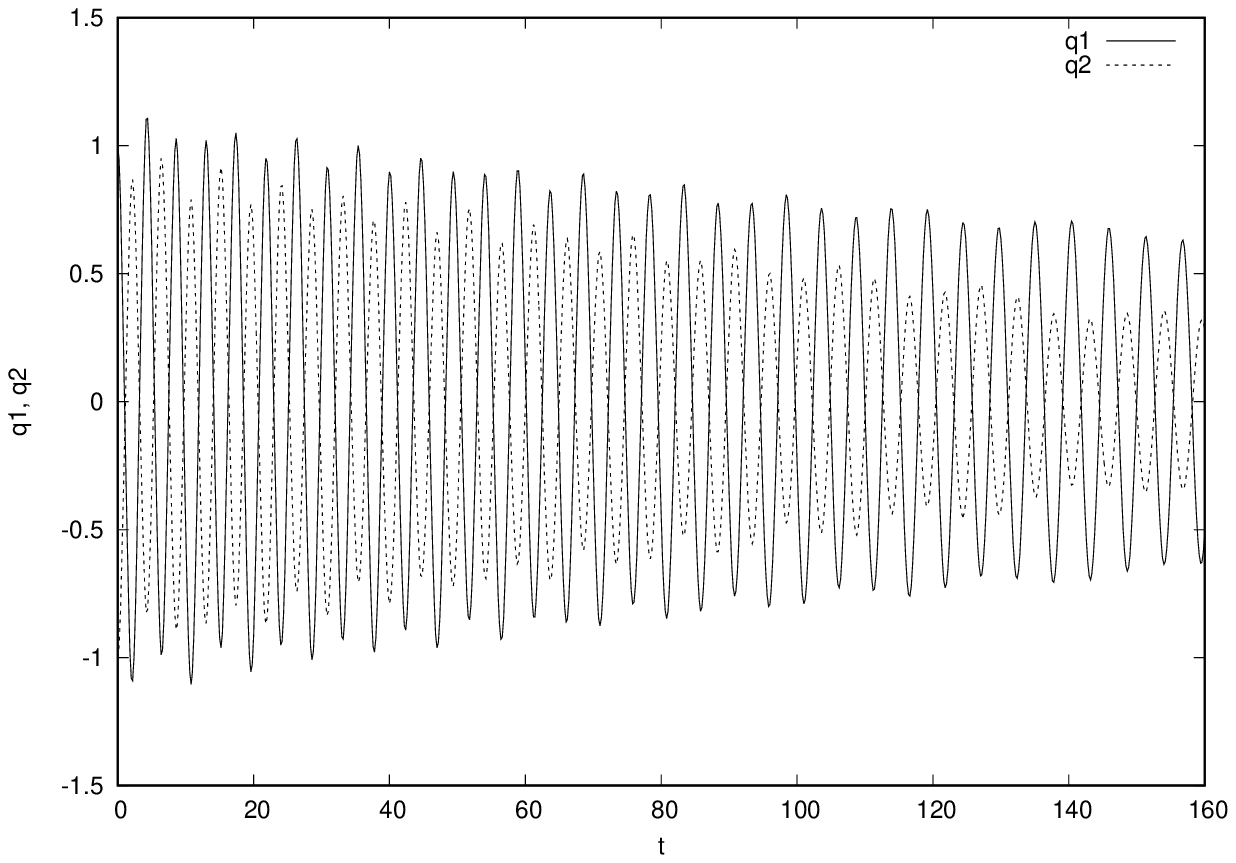}
\caption{$q_1(t), q_2(t)$ for $g=0.0$ (left) and $g=0.8$ (right).}
\end{figure}

\newpage

\section{Summary}
\setcounter{equation}{0}

Starting from a differential one-form with an integration factor $\lambda$
\begin{align}
\omega=pdq+dz-Kdt
\quad\Longrightarrow\quad 
\Omega=\lambda\omega,
\end{align}
we derived equations of motion driven by the contact Hamiltonian $K(p,q,z)$
\begin{align}
\dot{q}=K_p,\quad
\dot{p}=-K_q+pK_z,\quad 
\dot{z}=K-pK_p,\quad 
\dot{\lambda}=-\lambda K_z.
\label{EoMfinalS}
\end{align}

Then we introduced the contact Poisson bracket 
\begin{align}
\{A, B\}_\text{c}&=
\left(\frac{\partial A}{\partial q}\frac{\partial B}{\partial p}-
\frac{\partial B}{\partial q}\frac{\partial A}{\partial p}\right)
\nonumber \\
&\quad +\left[\frac{\partial A}{\partial z}
\left(\lambda\frac{\partial B}{\partial\lambda}-p\frac{\partial B}{\partial p}\right)-
\frac{\partial B}{\partial z}
\left(\lambda\frac{\partial A}{\partial\lambda}-p\frac{\partial A}{\partial p}\right)\right],
\end{align}
which enables us to write equation of motion for arbitrary quantity $A(p, q, \lambda, z)$ by
\begin{align}
\frac{dA}{dt}=\{A, K\}_\text{c}+A_zK.
\label{contactHamiltonS}
\end{align}
The last term represents a unique feature of the contact dynamics. 

The contact Lagrangian $J$ was introduced by
\begin{align}
J(q, \dot{q}, z)=p\dot{q}-K(p, q, z),\qquad
p=\frac{\partial J}{\partial \dot{q}},\quad
\dot{q}=\frac{\partial K}{\partial p},
\label{contactLagrangianS}
\end{align}
whose extended Euler-Lagrange equation is given by Herglotz equation
\begin{align}
\frac{d}{dt}\left(\frac{\partial J}{\partial\dot{q}}\right)-\frac{\partial J}{\partial q}
+\frac{\partial J}{\partial\dot{q}}\frac{\partial J}{\partial z}=0,
\end{align}
which is nothing but the second equation of \eqref{EoMfinalS},
\begin{align}
\frac{dp}{dt}=-K_q+pK_z.
\end{align}
Since the right hand side of \eqref{contactLagrangianS} becomes $pK_p-K=-\dot{z}$, which is the third equation 
of \eqref{EoMfinalS}, we have an equality $\dot{z}+J=0$. 
In other words, the contact variable $z$ has a meaning of (minus of) the action integral,
\begin{align}
dz=-Jdt \quad\Longrightarrow\quad
z=-\int J(q, \dot{q}, z)\ dt.
\end{align}
If we had started from one-form of $\omega=pdq-dz-Kdt$ instead of $\omega=pdq+dz-Kdt$, $z$ would be truly 
the action variable $z=\int J dt$, with plus sign in the right hand side of the above equation.

Finally symplectic integrator for the contact dynamics, which is named hybrid leap-frog method, was formulated 
for numerical computations, and applied to several examples such as damped harmonic oscillators, 
motion in double-well potential, and synchronization of two oscillators. Obtained results are quite reasonable. 

\section*{Acknowledgement}
The authors are very grateful for Dr. K.M. Aoki for valuable discussions and for teaching some references.

\end{document}